\documentclass[sigconf,screen,authorversion,nonacm]{acmart}
\AtBeginDocument{%
  }

\begin{document}
\copyrightyear{2026}
\acmYear{2026}
\setcopyright{cc}
\setcctype{by}
\acmConference[CHI EA '26]{Extended Abstracts of the 2026 CHI Conference on Human Factors in Computing Systems}{April 13--17, 2026}{Barcelona, Spain}
\acmBooktitle{Extended Abstracts of the 2026 CHI Conference on Human Factors in Computing Systems (CHI EA '26), April 13--17, 2026, Barcelona, Spain}
\acmDOI{10.1145/3772363.3798628}
\acmISBN{979-8-4007-2281-3/2026/04}

\author{Yao Xiao}
\authornote{Corresponding author.}
\affiliation{%
  \department{Dyson School of Design Engineering}
  \institution{Imperial College London}
  \city{London}
  \country{United Kingdom}}
\email{yxiao3@ic.ac.uk}

\author{Rafael A. Calvo}
\affiliation{%
  \department{Dyson School of Design Engineering}
  \institution{Imperial College London}
  \city{London}
  \country{United Kingdom}}
\email{r.calvo@imperial.ac.uk}

\title{AI as Relational Translator: Rethinking Belonging and Mutual Legibility in Cross-Cultural Contexts}


\begin{abstract}
  Against rising global loneliness, AI companions promise connection, yet accumulating evidence suggests that, for some users and contexts, intensive companion-style use can correlate with increased loneliness and reduced offline socialisation. This position paper challenges the dominant ``AI as companion'' paradigm by proposing a shift: from AI that simulates relationships \emph{with} humans to AI that supports relationships \emph{between} humans. We introduce \textit{Relational AI Translation}, positioning AI as cultural-relational infrastructure that scaffolds human connection across cultural, generational, and geographical divides. Using first-generation East Asian migrants as a theoretically productive critical case, we outline a multi-agent architecture instantiating three translation operations: emotion-intent decoding, contextual reframing, and relational scaffolding. We articulate design provocations around measurement, safety architecture, and the tension between technological intervention and structural justice, and explicitly frame success as graduation toward renewed human-to-human support rather than sustained engagement with the system. 
\end{abstract}



\begin{teaserfigure}
  \centering
  \includegraphics[width=\textwidth]{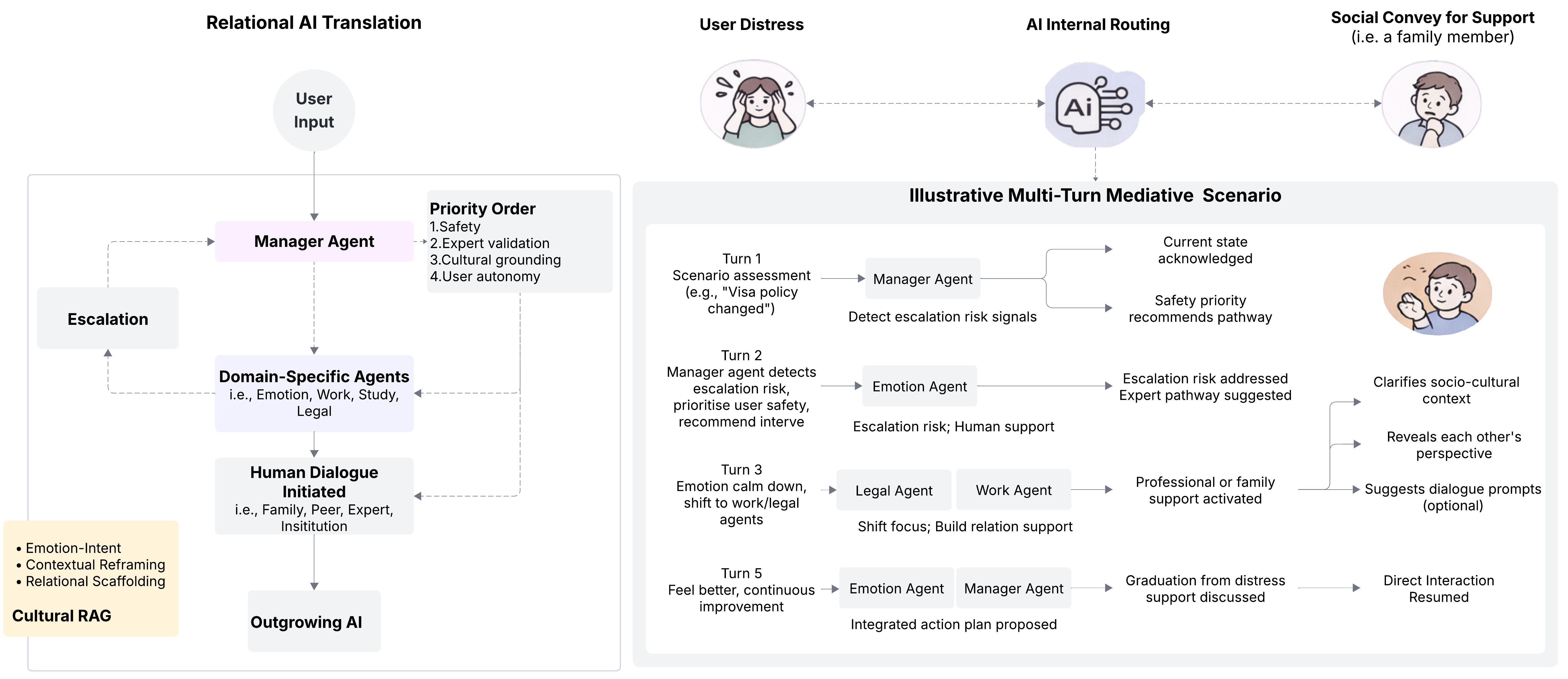}
  \caption{Relational AI Translation as multi-agent mediation and relational scaffolding.}
  \Description{A flow-based multi-turn mediation diagram. On the left, user input enters a manager agent that routes requests according to safety and escalation priorities. Domain-specific agents (emotion, work, legal) handle contextual needs, while escalation logic activates human dialogue when necessary. On the right, a multi-turn scenario demonstrates relational scaffolding: initial distress is acknowledged, escalation risk is assessed, socio-cultural context is clarified, implicit relational perspectives are surfaced, and dialogue prompts are optionally suggested. The process culminates in professional or family support activation and direct human interaction, followed by gradual withdrawal of AI mediation.}
  \label{fig:flowdia}
\end{teaserfigure}

\maketitle
\noindent\textbf{Publication note.}
Accepted for publication in \textit{CHI EA '26}.

\medskip
\noindent\textbf{Published version citation.}
Yao Xiao and Rafael A. Calvo. 2026. \textit{AI as Relational Translator: Rethinking Belonging and Mutual Legibility in Cross-Cultural Contexts}. In \textit{Extended Abstracts of the 2026 CHI Conference on Human Factors in Computing Systems (CHI EA '26)}, April 13--17, 2026, Barcelona, Spain. ACM, New York, NY, USA, 8 pages. \url{https://doi.org/10.1145/3772363.3798628}

\section{Introduction}

The \emph{companionship paradox} presents a fundamental challenge for AI and wellbeing. Companion systems promise always-available support, yet recent evidence suggests that longer or more intensive chatbot use can be associated with increased loneliness and reduced offline socialisation in some settings~\cite{Phang2025Investigating}. Meanwhile, users may report feeling seen and held, even as studies raise concerns about perceived empathy gaps, social deskilling, and the displacement of human-to-human relating~\cite{Kim2024AIInduced, Liu2025Illusion}.

We argue that this paradox reflects a category error. Dominant systems ask whether AI can \emph{be} a relationship, whereas a more productive design question concerns whether AI can mediate \emph{between} relationships.

We propose \emph{Relational AI Translation}, positioning AI not as a synthetic companion but as socio-technical infrastructure that helps humans negotiate relational meaning across contexts where mutual understanding may break down.
\subsection{Why a Critical Case: Migration as Relational Mismatch}
This work uses first-generation East Asian migrants in Western host societies as an analytically productive case study. In many East Asian contexts, cultural constructs such as “face” operate as co-constructed social currency grounded in relational recognition and social standing; losing face generates shame that reverberates across networks and shapes disclosure decisions. Therefore, such practices (e.g., \textit{mianzi} [CN], \textit{mentsu} [JP], \textit{chemyon} [KR]) can render psychological distress less legible as a cue for support seeking~\cite{Fei2023ChineseCulture,Sun2024Behavioural}. Relational constructs such as \textit{guanxi} (networked obligation) [CN] further structure interaction through linguistic indirectness, hierarchy awareness, and moralised emotion, where visible distress may be perceived as a burden.

For some first-generation migrants, geographic separation from familiar support networks may also widen this legibility gap, prompting exploration of indirect or low-burden mediation, including Conversational AI (CA).

However, prior research shows that mainstream CAs often encode Western individualistic assumptions~\cite{FenechBorg2025Cultural, Rao2025NormAd}, struggling in cultural contexts where relational needs are central~\cite{Pawar2024Presumed, Pawar2025Survey}. Existing work on cultural alignment~\cite{Lee2025IntoUnknown, Naidoo2025Culturally, Pawar2025Survey, Rao2025NormAd, Tao2024CulturalBias} frequently treats culture as static, overlooking how migrants’ cultural positioning shifts over time, producing model drift and misattunement.

\subsection{Analytical Scope}
This work focuses primarily on cultural-relational dynamics in interpersonal 
communication. Institutional power structures and structural precarity are 
acknowledged but not fully addressed. The framework is a bounded intervention 
for interpersonal dynamics, not comprehensive response to migration-related distress.

\section{The Mirage and the Illusion}

\subsection{Design Mirages and Mind Perception}

Abercrombie et al.~\cite{Abercrombie2023Mirages} describe \emph{mirages}, anthropomorphic design features such as first-person pronouns, realistic voices, and conversational disfluencies that encourage users to personify systems lacking human-like intentionality. Folk et al.~\cite{Folk2025CulturalVariation} further report that East Asian students expressed greater openness towards social chatbots than their European counterparts, mediated by a higher propensity to anthropomorphise technology.

Epley et al.’s theory~\cite{Epley2007OnSeeing} suggests that anthropomorphism increases when individuals seek social connection and predictability. Vulnerability, however, is unevenly distributed. Individuals experiencing loneliness or constrained social support may be more likely to attribute mind and care to interactive systems, increasing risks of disappointment, overreliance, or misplaced trust. These dynamics render mirage-driven designs ethically concerning in migration contexts~\cite{Berry1997Immigration, Lei2024Unpacking}. As discussed in the previous section, cultural misalignment in conversational agents can further exacerbate this risk when systems fail to respond appropriately to context-sensitive relational cues. 

\subsection{The Illusion of Understanding}

Large language models can perform well on certain theory-of-mind benchmarks while lacking higher-order affective understanding~\cite{Marchetti2025Artificial}. They replicate the outputs of mentalising without possessing underlying socio-relational capacities, producing what we term the \emph{illusion of understanding}. Users may feel understood by systems that cannot understand.

In multi-agent settings, this risk can compound. Errors may propagate across agents~\cite{Guo2024LLMSurvey}, producing cascading hallucinations in which an initial misinterpretation triggers chains of emotional misdirection. For wellbeing-adjacent applications, this underscores the need for explicit uncertainty handling, boundary-setting, and escalation pathways.
\subsection{AI–Human Relations as a Conditional Risk}

Even as affective modelling~\cite{Kosinski2024Evaluating} and cultural sensitivity advance, the core challenge is not technical approximation but socio-technical consequence: how might ``synthetic empathy'' restructure human-to-human relatability? 

Current discourse often questions whether synthetic empathy can meaningfully augment human connection, or if it risks displacing it. Indeed, prior work on mind perception~\cite{Gray2012MindPerception} suggests a potential displacement effect: the attribution of mental states to artificial agents has been correlated with a diminished perception of mind, namely, a dehumanisation effect toward human counterparts~\cite{Herak2020Pairing, Kim2024AIInduced}. Instead of treating this as inevitable, we treat it as a contingent risk. The emergence of such ``mind-perception trade-offs'' is likely modulated by interface design, situational context, and user vulnerability. In migratory or high-stakes social contexts, an over-reliance on AI-mediated channels may trigger subtle forms of dehumanisation, potentially diverting relational investment away from socially embedded others.

Guided by these tensions, our work adopts a design stance that moves beyond the pursuit of affective simulation. Instead of positioning the system as a surrogate for human interaction, we reimagine translation-oriented agents as relational scaffolding—infrastructural support designed to cultivate, rather than circumvent, genuine human belonging.

\section{Relational AI Translation: Conceptual Framework and Architecture}

In contrast to Claggett et al.’s Relational AI framework, which emphasises intergroup dynamics~\cite{Claggett2025Relational}, we focus on interpersonal misunderstandings in socio-culturally mismatched relationships. This work conceptualises \emph{Relational AI Translation} as a bounded socio-technical infrastructure designed to support human-to-human understanding across culture. It focuses on increasing \emph{mutual legibility} between people.

\begin{figure*}[t]
  \centering
  \includegraphics[width=0.82\textwidth]{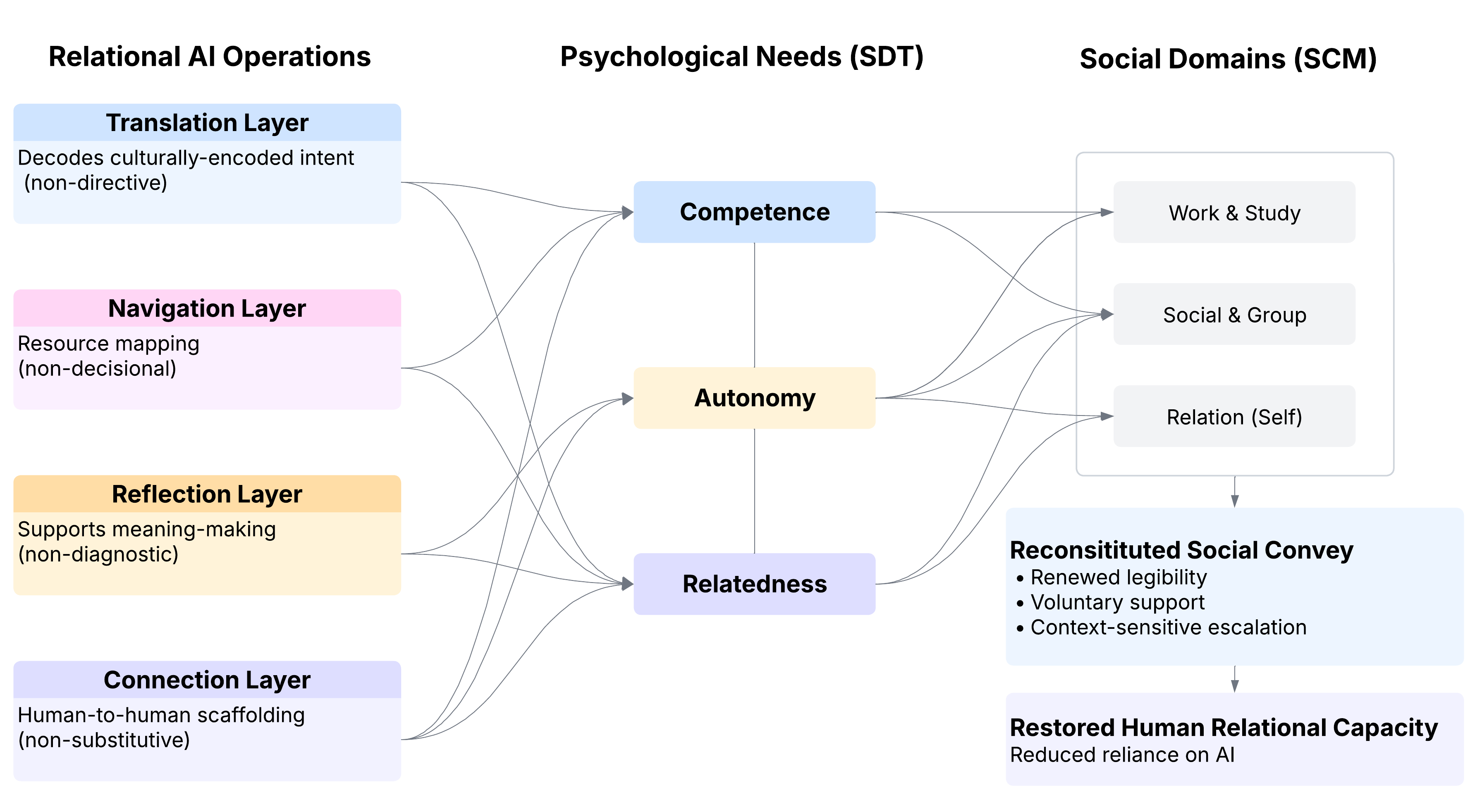}
  \caption{Operationalising Relational AI as bounded culture–relational infrastructure.}
  \Description{A three-column conceptual mapping diagram. The left column presents four Relational AI layers: Translation (decoding culturally encoded intent), Navigation (resource mapping), Reflection (meaning-making support), and Connection (human-to-human scaffolding), all marked as non-directive and non-substitutive. The centre column maps these layers onto psychological needs from Self-Determination Theory: competence, autonomy, and relatedness. The right column connects these needs to social domains: work and study, social and group contexts, and relation to self. The bottom shows a reconstituted social convoy enabling renewed legibility, voluntary support, and context-sensitive escalation, culminating in restored human relational capacity and reduced reliance on AI.}
  \label{fig:rait-theory-framework}
\end{figure*}

\subsection{Theoretical Integration}
Relational AI Translation integrates Self-Determination Theory (SDT)~\cite{Calvo2014Positive,Ryan2000SelfDetermination} with the Social Convoy Model (SCM)~\cite{Kahn1980Convoys} to address the sociotechnical challenges of migration. SDT provides a lens for mapping cultural contexts onto psychological needs (autonomy, competence, relatedness), while SCM foregrounds the erosion and restructuring of support networks during relocation~\cite{Lei2024Unpacking}. 

Figure~\ref{fig:rait-theory-framework} maps four non-substitutive AI layers: translation, navigation, reflection, and connection, onto Self-Determination Theory (competence, autonomy, relatedness) across social domains. Rather than substituting relationships, the system supports renewed legibility, voluntary activation, and context-sensitive escalation within the social convoy. Here, success is measured not by AI engagement, but by restored human relational capacity and reduced reliance on AI. We propose an architecture that scaffolds cross-contextual understanding and bridges users to human resources, explicitly rejecting AI as a pseudo-member.

\subsection{Translation Operations}
Relational AI Translation operationalises its approach through three system-level operations.

\emph{Emotion-intent decoding} attends to culturally situated meanings. For example, when a user says, ``I'm fine'' (e.g., \textit{meiguanxi} [CN], \textit{daijobu} [JP], \textit{gwaenchana} [KR]), the system does not assume the statement reflects absence of concern. This interactional stance prioritises care and confirmation over cultural assumption.

\emph{Contextual reframing} surfaces multiple relational perspectives to support mutual understanding. For example, intergenerational conflict can be examined through both autonomy-oriented and obligation-oriented lenses, inviting perspective-taking and relational empathy without pathologising either position.

\emph{Relational scaffolding} supports human-to-human communication across contexts. The system assists with culturally relevant communication techniques. Where appropriate, it also connects users to relevant human and community-based resources to support these interactions, strengthening mutual legibility between people.

\subsection{Technical Instantiation (Conceptual Prototype)}

Relational AI Translation is instantiated through a multi-agent architecture 
~\cite{Guo2024LLMSurvey, Park2023Generative} organised around recurring relational domains identified in migration and wellbeing literature. As illustrated in Figure~\ref{fig:flowdia}, the system processes user distress through a hierarchical coordination of manager and domain-specific agents. This process follows a rigorous priority order: safety, expert validation, cultural grounding, and user autonomy, to ensure ethical alignment. Escalation logic enforces system boundaries by activating human support when risks are detected. Across multiple turns, translation operations clarify socio-cultural contexts, surface implicit relational perspectives, and scaffold dialogue between humans, with the ultimate goal of restoring direct human interaction and gradually reducing AI mediation.

A Manager agent orchestrates system behaviour and coordinates domain-specific agents (e.g., Work, Study, Emotion, Legal) operating within bounded scopes. A Cultural RAG layer retrieves co-designed cultural materials developed through participatory and community-engaged methods.

When agent outputs diverge, coordination follows a defined priority order: (1) safety and crisis escalation, (2) expert validation in high-stakes cases, (3) cultural grounding checks, and (4) user autonomy when multiple low-risk options remain. This architecture separates routing, cultural grounding, and domain reasoning.

\section{Design Provocations}
The following provocations articulate core tensions raised by Relational AI 
Translation, focusing on evaluative measurements, cultural risk, dependency prevention, and structural limits.
\subsection{Separating Cultural Translation from Emotional Support}

\emph{What if observed wellbeing improvements stem from empathetic warmth rather 
than cultural translation?}

Wellbeing improvements may arise from general empathetic warmth rather than translation-specific mechanisms. If culturally attuned mediation is combined with emotional affirmation (e.g., ``I hear that's difficult''), observed gains risk being misattributed. We thus propose component-level evaluation comparing translation operations against a supportive baseline without cultural mediation. Mixed methods including experience sampling, behavioural traces, and conversation analysis can help isolate specific effects.

\subsection{Protective Ambiguity and the Dual Risks of Translation}

\emph{What if translation disrupts protective ambiguity or misfires altogether?}

High-context communication often relies on strategic indirectness, silence, and face-sensitive restraint. Implicit understanding is not a deficit but a relational strategy. Over-explication risks imposing Western preferences for explicitness, intensifying vulnerability, and producing ``synthetic harmony'' that reinforces AI reliance rather than supporting graduation to human connection. Treating cultural categories (e.g., ``East Asian'') as fixed profiles risks culture essentialism and stereotyping.

Translation thus carries a dual risk: \textit{over-translation} (eroding protective ambiguity) and \textit{mis-translation} (incorrectly inferring culturally situated meaning). 

To mitigate these risks, cultural interpretations are framed as provisional hypotheses rather than determinate facts. The system surfaces uncertainty, invites user confirmation, and treats cultural lenses as adjustable through interaction rather than from static demographic profiles. Translation remains optional and user-directed, and ambiguity is recognised as a legitimate relational choice.
\subsection{Safety Boundaries, Activation, and AI Authority}

\emph{What if systems treated ``I cannot help'' as designed capability, and activation prompts as invitations rather than prescriptions?}

In wellbeing-adjacent contexts, relational mediation may encounter crisis (suicidal 
ideation), discrimination, or legal precarity that exceed AI's appropriate scope. 
Simultaneously, Behavioural Activation~\cite{Sun2024Behavioural}, engaging in valued 
activities to improve mood, while valuable for low-intensity wellbeing support, 
raises risks in migration contexts where users face relational constraints 
(obligations to care for parents, fears of losing face) that AI cannot navigate 
authoritatively.

The system therefore foregrounds calibrated limitation and transparent boundary-setting. Activation prompts are framed as invitations rather than prescriptions, preserving user agency. Escalation to human or institutional support is treated as responsible mediation, not breakdown. Transparency about limits becomes part of relational scaffolding itself.

\subsection{Embodiment, Dependency, and Structural Limits}

\emph{What if the system creates permanent dependency or obscures structural 
injustice?}

As AI systems become more embodied or emotionally expressive, risks of dependency and authority attribution intensify. To prevent synthetic harmony, the system inverts engagement-optimisation logics: it teaches relational principles rather than providing solutions, fades scaffolding over time, and encourages independent problem-solving. Success is measured by decreased AI reliance alongside increased offline connection.

Relational distress may also reflect power asymmetries, value conflicts, or structural injustices, such as visa precarity, workplace discrimination, or inadequate healthcare, that translation alone cannot resolve. The system therefore foregrounds user welfare by documenting structural barriers and connecting users to community, legal, and social resources. Success is defined by users gradually outgrowing the system as their social convoys rebuild.

\section{Limitations and Future Work}

Relational AI Translation positions AI as bounded cultural-relational infrastructure that scaffolds human-to-human understanding. This work is in an early conceptual stage. Open challenges include model drift as cultural contexts evolve, autonomy–safety trade-offs, and limitations in memory architecture and accessibility. Mutual legibility is not universally valued, and cultural reasoning may misinterpret or essentialise despite safeguards. Finally, the framework addresses interpersonal mediation rather than institutional reform. Many relational tensions reflect structural injustice or value conflicts beyond the focus of the current scope.

Future work will extend this framework through empirical participatory studies~\cite{Calvo2023Assessing,Hedditch2025Building,Klassen2024BlackToFuture}. Evaluation will examine communication self-efficacy, perceived comprehension, and changes in offline engagement, with graduation defined as reduced AI reliance. Cross-cultural and longitudinal inquiry will test generalisability and sustained impact.

\section{Conclusion}
AI systems are frequently framed as emotional companions. We propose an alternative orientation: AI as a cultural-relational mediator, with success defined by restored human relational capacity and reduced reliance on the system.
\begin{acks}
YX is supported by the Hans Rausing Scholarship at Imperial College London. RAC acknowledges funding from the Leverhulme Centre for the Future of Intelligence.
\end{acks}

\bibliographystyle{ACM-Reference-Format}
\bibliography{sample-base}

\end{document}